\documentclass{article}
\usepackage{spconf,amsmath,graphicx}
\usepackage[hidelinks]{hyperref}

\title{Fast and small footprint Hybrid HMM-HiFiGAN based system for  speech synthesis in Indian languages}
\name{Sudhanshu Srivastava$^{1}$, Ishika Gupta$^1$, Anusha Prakash$^{2}$, Jom Kuriakose$^1$, Hema A. Murthy$^1$ }
\address{ 
 $^1$Department of Computer Science and Engineering, Indian Institute of Technology, Madras \\
 $^2$Department of Electrical Engineering, Indian Institute of Technology Madras \\
{\tt\small srivastava.rishabh4321@gmail.com, ishika@cse.iitm.ac.in, anushaprakash90@gmail.com} \\ {\tt\small jom@cse.iitm.ac.in,  hema@cse.iitm.ac.in}}
\begin{document}
%
\maketitle
\begin{abstract}
Hidden-Markov-model (HMM) based text-to-speech (HTS) offers flexibility in speaking styles along with fast training and synthesis while being computationally less intense. HTS performs well even in low-resource scenarios.   The primary drawback is that the voice quality is poor compared to that of E2E systems.   
A hybrid approach combining HMM-based feature generation and neural-network-based HiFi-GAN vocoder to improve HTS synthesis quality is proposed. HTS is trained on high-resolution mel-spectrograms instead of conventional mel generalized coefficients (MGC), and the output mel-spectrogram corresponding to the input text is used in a HiFi-GAN vocoder trained on Indic languages,  to produce naturalness that is equivalent to that of E2E systems, as evidenced from the DMOS and PC tests.
\end{abstract}
\vspace{-.1cm}
\vspace{-.4cm}
\section{Introduction}
\label{intro}
State-of-the-art E2E systems produce high-quality speech but incur significant compute resources and data, and also have a large footprint in terms of models. Unit selection speech synthesis systems are fast and produce high-quality speech but, joins can be abrupt, and have large footprints.  The objective of this paper is to improve the quality of hidden Markov model (HMM) based speech synthesis systems (HTS) \cite{tokuda,simon}.  The advantage of HTS-based speech synthesis systems is their small footprint and robustness in low-resource scenarios.   HTS requires a STRAIGHT vocoder \cite{STRAIGHT} (proprietary) to produce high-quality speech. In this paper, we try to marry the technology of HTS and the HiFi-GAN vocoder to produce high-quality speech at much lower computational and memory costs. 
Sequence-to-sequence (S2S) auto-regressive models, for example, Tacotron2 \cite{tacotron2} and Transformer TTS \cite{transformertts} models, combined with vocoders, such as WaveNet \cite{wavenet} and Waveglow \cite{waveglow}, produce good quality speech. Despite this significant improvement, there is a trade-off between quality and computational complexity. The high quality is limited to read speech, in which sentences are grammatically correct. Since E2E uses the entire sentence context, it is unable to scale up for conversational type text, and often has word skips and other artifacts \cite{BGM}. Although non-auto regressive  Fastspeech \cite{fs2} seldom has skips, nevertheless, they perform poorly in low-resource scenarios.
Subword models in HTS-based synthesis correspond to a pentaphone. Owing to tree-based clustering out-of-vocabulary words are also synthesizable. Nevertheless, synthesized speech in HTS-based systems is poor.
 In our earlier work \cite{is2020} it was shown that the quality of the synthesized output is primarily dependent on the fidelity of the mel spectrograms. 
A hybrid TTS approach combines two or more TTS frameworks. The current work is inspired by \cite{is2020}, where speech synthesized by HTS was converted to mel spectrograms required by the waveglow vocoder. In \cite{is2020}, mel spectrograms are extracted from the HTS-generated audio (for training text) and the ground-truth (GT) audio.
For each mel-filter coefficient,  one-dimensional histograms are estimated for the source HTS audio, and ground truth original audio. Histogram equalization is performed between the source and target histograms. During testing, the speech is first synthesized using HTS and converted to a sequence of mel-filter coefficients.  A lookup table is used to replace the HTS-generated mel-filter coefficients with that of the ground truth mel-filter coefficients. The speech is then synthesized using the waveglow vocoder. 
An attempt was also made to train the waveglow vocoder directly from the HTS-generated MGCs, but the output suffered in terms of timbre.  The HTS-generated MGCs lack the spectral resolution required.   In the current work, we first train the  HTS system using high-resolution mel-filters.  We replace the vocoder with the HiFi-GAN vocoder since it offers greater flexibility in terms of parameter choices\footnote{HiFi-GAN is a generative adversarial network (GAN) \cite{GAN} based vocoder with a synthesis quality better than autoregressive models such as WaveNet, and since it employs a non-autoregressive architecture, the inferencing is comparatively fast and does not require a GPU.}.   \\
  Recent hybrid systems include \cite{USS_hybrid,dypat,unitnet}. In \cite{USS_hybrid}  the best-suited polysyllable segments across USS and E2E synthesized speech are used.  Appropriate segments are chosen from the proposed hybrid system using either the USS units or the E2E units. \cite{unitnet} and \cite{uss_s2s} combine traditional methods with neural-network-based architectures like Tacotron2. \cite{uss_s2s} represents phones as vectors output by Tacotron encoder and uses Euclidean distance between them to perform unit selection, while \cite{unitnet} synthesizes extra phone-level units in the E2E context depending on selection cost. \cite{dypat} 
  uses a hidden semi-Markov model (HSMM) for latent representation of alignments to reduce the exposure bias, i.e., the mismatch between training (teacher-forcing) and inference (free-running). \cite{conversational_is22} is a recent work on conversational speech synthesis, where enhancing training data with conversational speech improves the prosody of conversational speech.
\\
 In the current work, we propose an approach to combine HTS and HiFi-GAN. The baseline systems consist of conventional HEQ and Fastspeech2 with a HiFi-GAN vocoder.
The HEQ model is similar to our earlier work \cite{is2020}, except that the waveglow vocoder is replaced by the HiFi-GAN vocoder. In the proposed approach, the HTS system is trained on mel-spectrograms. 
The model offers GPU-free inferencing with a total footprint of $\sim$ 72 MB (including the vocoder), making it deployable. It is observed that the model trained using the proposed approach generates good-quality speech even with 1 hour of labeled data.
The rest of the paper is organized as follows: Section \ref{related} reviews the related works. The proposed approaches are presented in Section \ref{sec:proposed}. Experiments and results are discussed in Section \ref{expts}. The work is concluded in Section \ref{conclusion}.
\vspace{-.4cm}
\section{Related Works}
\label{related}
\vspace{-.1cm}
\subsection{HTS System}
\label{HTS}
\vspace{-.1cm}
HTS is based on the source-filter model of speech \cite{sourcefilter}, wherein spectral and excitation parameters are convolved to produce speech. Mel-generalized cepstrums (MGCs) and log f0 (lf0, l$\delta$ f0, l$\delta \delta$ f0. f0 -- fundamental frequency) are the spectral and excitation parameters, respectively, which are extracted from the audio files. The HTS pipeline consists of a training phase and a testing phase.
\vspace{-.2cm}
\subsubsection{Training Phase of the text to speech synthesis system}
\vspace{-.2cm}
HMM-based speech synthesis uses either a small amount of labeled data to bootstrap the phone models or uses a flat start where it is assumed that all phones in a given utterance (as indicated by the text) are of equal duration.  Iterative embedded reestimation is performed to correct the boundaries and reestimate phone models. HTS is trained on labeled/aligned audio and text data. HMMs are trained to model context-dependent pentaphone ( 2 preceding + 2 succeeding + current phone) units. Decision tree-based clustering is performed to model a wide range of contextual labels thus making the system robust to unseen sequences.
\vspace{-.4cm}
\subsubsection{Speech generation using the models generated}
\vspace{-.26cm}
The test text is parsed into labels using the unified parser \cite{arunTSD16}. Then the HMMs corresponding to these labels are concatenated.  The spectral and excitation parameters are generated by HTS.  Mel Log Spectrum Approximation (MLSA) \cite{MLSA}, STRAIGHT \cite{STRAIGHT} or other vocoders can be used.
\vspace{-.2cm}
\vspace{-.18cm}
\subsection{HiFi-GAN}
\label{hifidetails}
\vspace{-.1cm}
HiFi-GAN \cite{hifiGAN} is a non-autoregressive state-of-the-art neural vocoder. It has one generator and two discriminators, which are trained adversarially. Along with the multi-scale discriminator proposed in MelGAN \cite{MelGAN} to model consecutive and long-term dependency, it uses a multi-period discriminator consisting of several sub-discriminators to handle a portion of the periodic signal of input audio. Moreover, a new mel-spectrogram loss (L1 distance of mel-spectrograms of original and generated waveforms) is added to the GAN objective. 
\vspace{-0.8cm}
\section{Proposed Approach}
\label{sec:proposed}
\begin{figure}[!h]
             \centering
             \includegraphics[width=\linewidth, height=5cm]{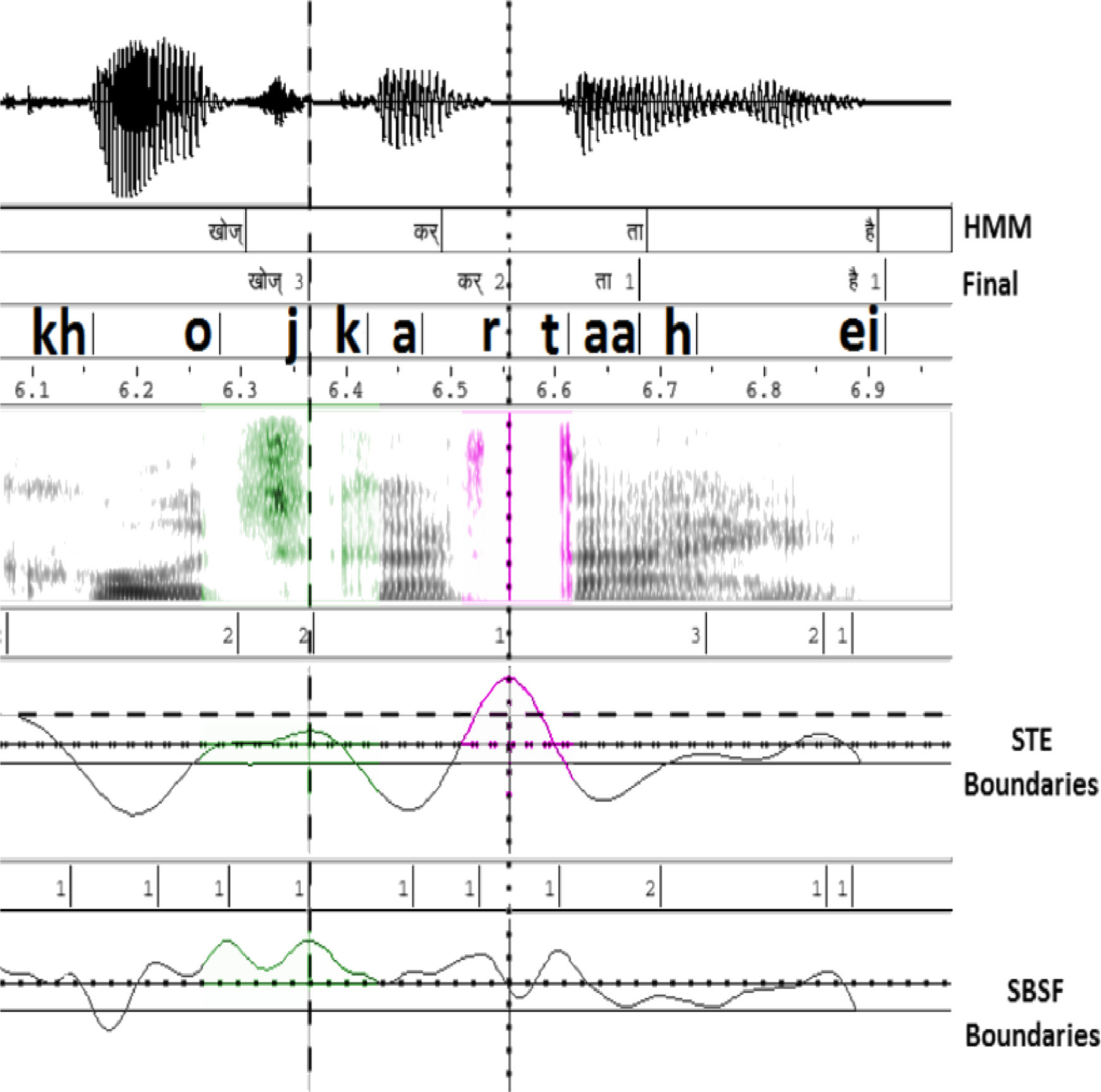}
             \caption{Hybrid Segmentation. The dots in pink show the STE and the green dots show the boundaries being corrected by SBSF. This figure is reproduced from \cite{hs_gd} with permission from the authors.}
             \label{HS_diagram}
             \vspace{-0.3cm}
 \end{figure}

\begin{figure}[!h]
             \centering
             \includegraphics[width=0.9\linewidth, height=5cm]{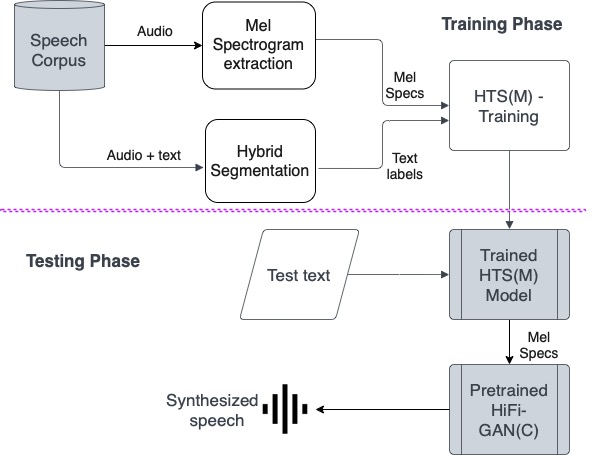}
             \caption{Flowchart of training and testing phases in proposed system.}
             \label{sys2}
             \vspace{-0.5cm}
 \end{figure}

\vspace{-.2cm}
In the proposed work we differ from existing approaches on two fronts: a)    While the flat start is used for initial segmentation, the boundaries are corrected using a language-agnostic signal processing approach which primarily uses the acoustic properties of syllables \cite{dnnseg,gd}.   The UTF-8 text is converted to a sequence of labels (both phone and syllable) using \cite{arunTSD16}.  b) Conventional HTS training uses Mel-Generalised Cepstral coefficients, whereas in the proposed approach we use high-resolution mel-filter bank coefficients. 
Group delay (GD) based segmentation exploits the property that each syllable has as its nucleus a high energy region corresponding to that of a vowel, an onset and a coda consisting of a sequence of consonants (C*VC*).   HMMs do not model boundaries well, but they give the correct number of boundaries when force aligned with text. Since GD segmentation is primarily based on signal processing, additional spurious boundaries can result.  The HMM boundaries are corrected using the accurate boundaries produced by GD processing.
The GD based processing does not give accurate boundaries for fricatives and affricates. Fricative and affricate boundaries are corrected using sub-based spectral flux (SBSF)\cite{gd}.   
This leads to segmentation of the speech signal into a sequence of syllables.  Embedded reestimation is performed at the syllable level.  This leads to  accurate phone boundaries, as the duration of the syllable is on an average about 130ms.  On the other hand, when embedded reestimation performed at the sentence level the duration can be as long as 15s, leading to poor phone boundaries as indicated in the Figure \ref{HS_diagram}.
 Figure \ref{HS_diagram} shows a segment of speech and boundaries obtained using this approach \cite{hs_gd}.  In Figure \ref{HS_diagram}, the boundary between syllables \textit{khoj} and \textit{kar} is corrected using GD of SBSF, and the boundary between syllables \textit{kar} and \textit{taa} is corrected using GD.
The boundary between two syllables is corrected using short-term-energy (STE) if the end phone of the first syllable is not a fricative or nasal and the beginning phone of the second syllable is not a  fricative, affricate, nasal, or semi-vowel. The boundary between two syllables is corrected using SBSF if the end of the first syllable or beginning of the second syllable, but not both, is a fricative or an affricate.
The vocoders used in HTS include MLSA \cite{MLSA} and STRAIGHT \cite{STRAIGHT}, which use the MGCs and pitch generated to synthesize speech. 
In the current work, we replace these vocoders with the vocoders used in E2E synthesizers. In the E2E framework, vocoders include wavenet \cite{wavenet}, waveglow \cite{waveglow}, HifiGAN, CARGAN \cite{cargan}.  In all of these, it is to be noted that the resolution of the spectrum is much higher than what is used in conventional HTS systems (34 mel-filters are used conventionally in HTS systems). Wavenet, waveglow, and HiFi-GAN use 80 mel filters by default. We, therefore, train the HTS system with 80 and 120 mel-filters.
\vspace{-.1cm}
Fig.\ref{sys2} shows a flowchart of the training and synthesis phases of the proposed system. The modified HTS (HTS on mel-spectrograms) and the conventional HiFi-GAN are referred to as HTS (M) and HiFi-GAN (C), respectively. During synthesis, the test sentence is passed through the HTS (M) system, which generates the corresponding mel-spectrogram.  The mel-spectrogram is fed to the HiFi-GAN (C) vocoder to produce the final speech output. In this approach, we also perform experiments with 80 and 120 mel-filters.

\vspace{-.3cm}
\section{Experiments and Results}
\label{expts}
\vspace{-.2cm}
This section discusses the datasets used and the various experiments carried out along with their results.
\vspace{-.3cm}
\subsection{Datasets used}
\vspace{-.2cm}
The proposed systems are tested in two different languages, Hindi and Kannada.
The HTS and HiFi-GAN modules are trained on 8.5 hours of Hindi-male and Kannada-male datasets respectively. Datasets are obtained from IndicTTS database \cite{indicTTS}. IndicTTS is an open-source TTS database recorded at 48kHz in a studio environment.
For the testing, we use translated transcripts from the Study Webs of Active–Learning for Young Aspiring Minds (SWAYAM) platform \cite{swayam}. The original SWAYAM lectures are in English. A state-of-the-art automatic-speech-recognition (ASR) \cite{ASRIITM} generates the transcription which is translated to Hindi and Kannada using a machine translation (MT) system \cite{multiswayam}. The generated transcripts are conversational in nature. In addition, we also synthesize general domain sentences for testing the scalability of the proposed systems.
\vspace{-.3cm}
\subsection{Experiments}
\vspace{-.2cm}
The conventional HTS+HEQ system (as described in \cite{is2020} is considered the baseline system. A Fastspeech system with a HiFi-GAN vocoder is also trained for comparison in terms of speech quality. The Fastspeech model is trained using ESPnet toolkit \cite{espnet-tts}. The alignments are obtained from tacotron2\cite{tacotron2} teacher model. We also evaluate the quality of the proposed system in a low-resource scenario, wherein only 1 hour of data is used for training. In this low-resource scenario, an HTS can be trained, albeit with degradation in synthesis quality. However, the Fastspeech  model did not train well, due to the lack of data.
Conventionally, mel-spectrograms are generated with 80 mel-filters. While we train with 120 mel-filters since we do not use the pitch generated by HTS in the HifiGAN vocoder.
\vspace{-.3cm}
\subsection{Evaluation}
\vspace{-.2cm}
A degradation mean opinion scores (DMOS) test is conducted to evaluate the performance of the various systems. In the DMOS test, native speakers of the language listen to the synthesized audio and rate utterances on a scale of 1-5, with 1 being poor and 5 being human-like. GT utterances are included for reference. The utterances are presented to the evaluators in random order. The final score is reported after normalization with respect to the scores of the GT utterances.
A pairwise comparison test (PC test) was also conducted between the proposed hybrid system and the HEQ hybrid system. In this measure, the evaluators listen to the same synthesized utterance generated by both systems and indicate their score as their preference.
Table 1 presents the DMOS results for systems in Hindi and Kannada. A total of 20 utterances (8 from each system + 4 GT) were evaluated by 15 (Hindi) and 10 (Kannada) listeners in the DMOS test. The last row displays the model footprint without including the vocoder size. HEQ model footprint is $\sim$ 17.5 MB. The proposed system outperforms FS for Kannada. Dravidian languages have significant agglutination, which can lead to poor alignments,  which can affect synthesis performance.   
Table 2 presents the PC results for systems in Hindi and Kannada.  15 (Hindi) and 10 (Kannada) native speakers evaluated a set of 10 pairs of utterances synthesized by the above-mentioned systems. This [\href{https://www.iitm.ac.in/donlab/preview/ICASSP2023/index.html}{\textit{\textbf{\underline{link}}}}] contains samples of 1-hour HTS (mel-spec), 8.5 hours 80 mel filterbanks HTS and conventional HTS generated utterances along with above-stated models.
\vspace{-.3cm}
\begin{table}[]
\label{dmos}
\caption{DMOS evaluation results comparing the synthesized utterances generated by Fastspeech (FS) and proposed approach}
\centering
\begin{tabular}{|l|l|l|}
\hline
Language                               & FS  & Proposed      \\ \hline
Hindi                                  & 4.32 & \textbf{3.99} \\ \hline
Kannada                                & 3.01 & \textbf{3.22} \\ \hline
Footprint size (MBs) (without vocoder) & 146  & \textbf{15.9}          \\ \hline
\end{tabular}
\vspace{-0.5cm}
\end{table}
\begin{table}[]
\centering
\caption{PC test results: HEQ system vs proposed hybrid system (preference in \%)}
\begin{tabular}{|l|l|l|l|}
\hline
Language & HEQ & Proposed   & Equal \\ \hline
Hindi    & 17.5   & \textbf{60} & 22.5     \\ \hline
Kannada  & 20   & \textbf{63.33} & 16.66     \\ \hline
\end{tabular}
\vspace{-0.5cm}
\end{table}
\vspace{-0.25cm}
\vspace{0.05cm}
\subsection{Results and discussions}
\vspace{-0.2cm}
 \begin{figure}[!h]
             \centering
             \includegraphics[width=0.9\linewidth]{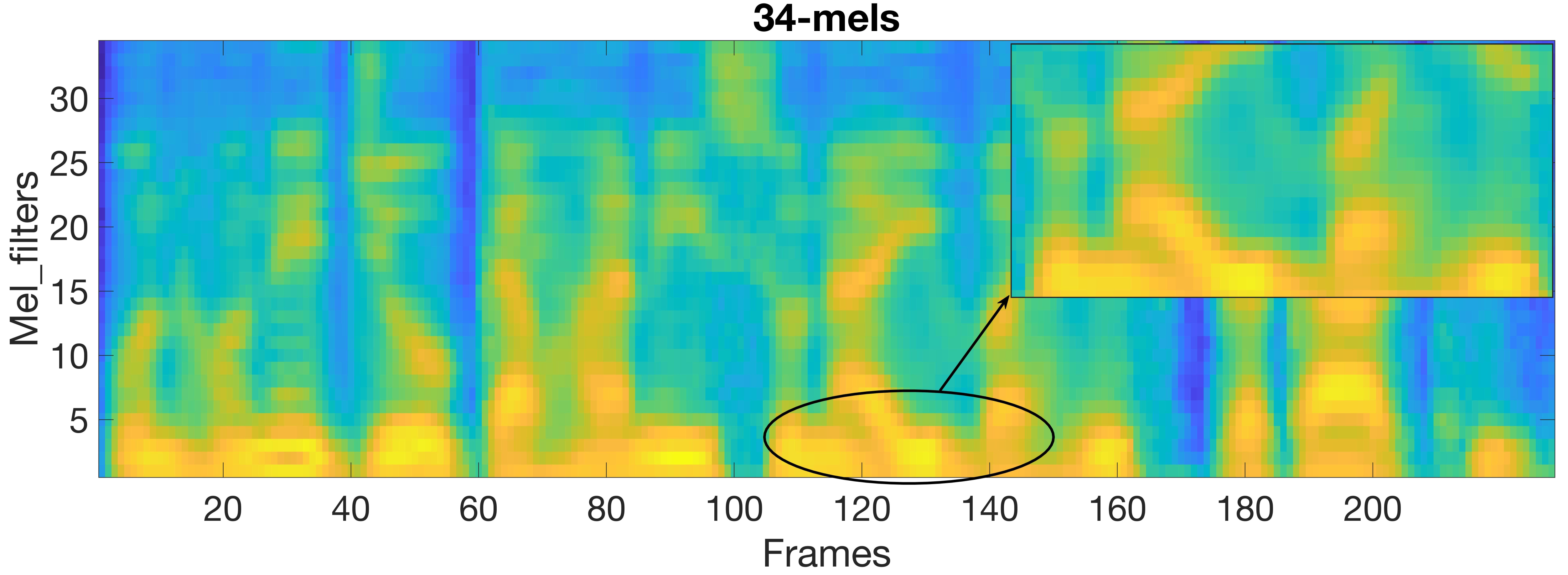}
             \centerline{(a) 34 mel filterbanks}
             \includegraphics[width=0.9\linewidth]{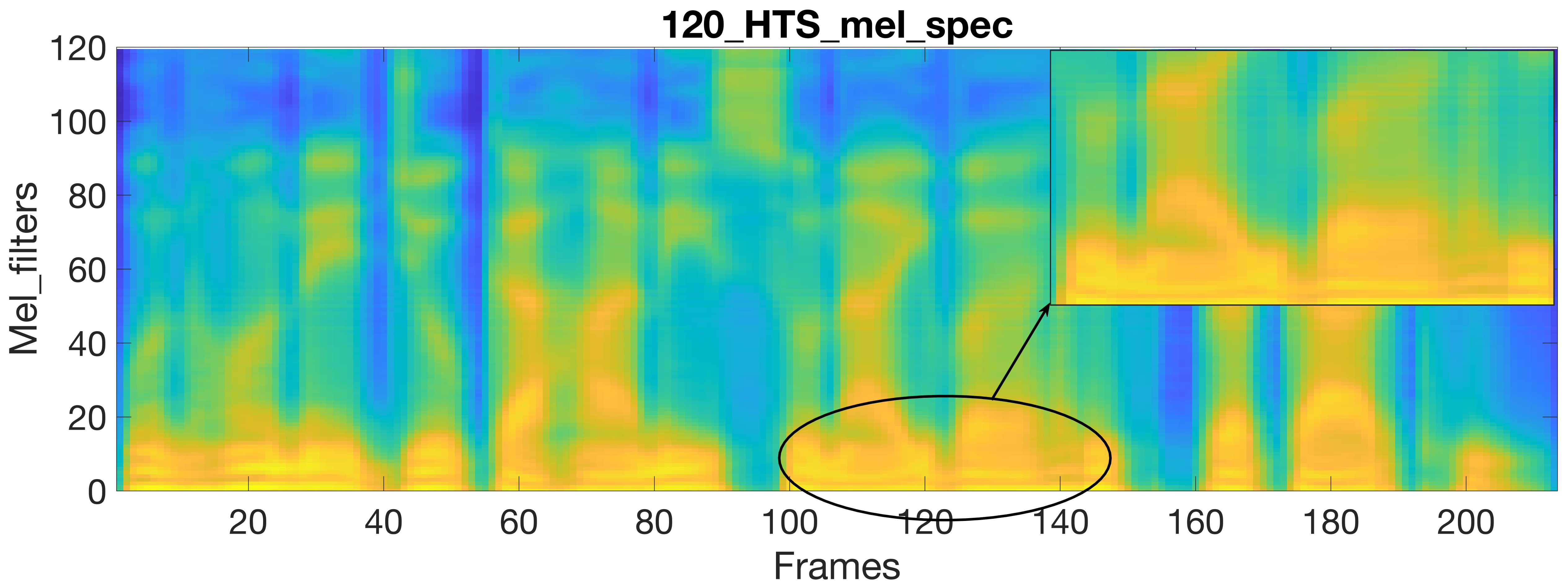}
             \centerline{(b) 120 mel filterbanks (Proposed Approach)}
             \includegraphics[width=0.9\linewidth, height=2.6cm]{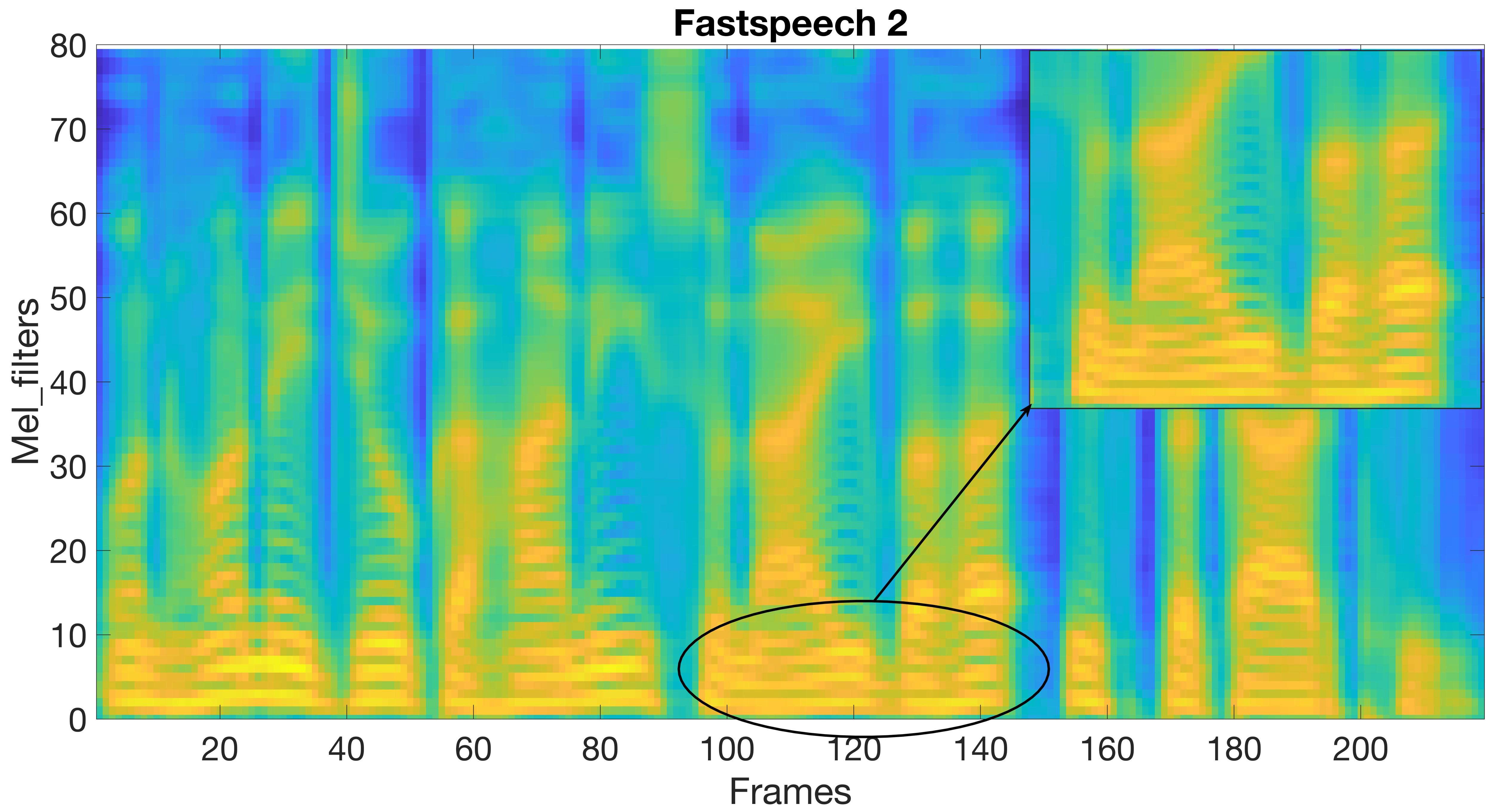}
             \centerline{(c) 80 mel filterbanks (Conventional Fastspeech)}
             \caption{Mel spectrograms across different systems.}
             \label{mels}
             \vspace{-0.45cm}
 \end{figure}
From Tables 1 and 2, it is seen that the proposed system performs better than the baseline HEQ system and is comparable to Fastspeech.  However, in the low-resource scenario, the Fastspeech model fails to train, while the proposed approach produces reasonably intelligible speech. [\href{https://www.iitm.ac.in/donlab/preview/ICASSP2023/index.html}{\textit{\textbf{\underline{link}}}}] shows sample 1-hour utterances.
Moreover, in a low-resource environment, if we compare the mel-spectrogram extraction time (on a single CPU core) for a sample sentence of 8 words, the amount of CPU time (user mode) spent by FS is 3.591s which is considerably high compared to 0.488s of the HTS system. 
Training of Hifi-GAN using CPU only is slow.  However, the inferencing is GPU-independent. This [\href{https://drive.google.com/drive/folders/13wJvcNst40cxZ5qKdaHqmGYHUguuswEn?usp=sharing}{\textit{\textbf{\underline{link}}}}]  contains vocoders trained from scratch on the IndicTTS database. Figure \ref{mels} shows the mel-spectrograms across various systems (34 mel filters as used in conventional HTS, \ref{mels}(a)). 
The spectrogram generated in Fig. \ref{mels}(b) is closer to that produced by fastspeech2.
\vspace{-.6cm}
\vspace{.2cm}
\section{Conclusions}
\label{conclusion}
\vspace{-0.2cm}
The proposed system (HTS+HiFi-GAN) capitalizes on the benefits of HMM-based speech synthesis, where out-of-vocabulary words can be modeled, and HiFi-GAN vocoder, which is capable of fast inferencing with good quality synthesis. The most important advantage of the proposed systems is the small footprint size. The mel-spectrogram generator in the proposed system is $\sim$16 MBs, in comparison to a neural network-based Fastspeech model ($>$100 MB to GBs). The HiFi-GAN vocoder has a footprint of $\sim$ 56 MB and the inference is GPU-free.  
Furthermore, the model gives intelligible results even with 1 hour of transcribed data used for training an HTS system, whereas E2E systems do not train.  Such a system is attractive primarily from the point of view of its small footprint and fast inferencing and, can be ported to smartphones.
\vspace{-.62cm}
\begin{center}
    \section*{Acknowledgment}
\end{center}
\vspace{-.25cm}
This work was carried out as a part of the project, ``Speech Technologies in Indian Languages ''(SP21221960CSMEIT0-\\03119) funded by the Ministry of Electronics and Information Technology (MeitY).
\\
\bibliographystyle{IEEEbib}
\bibliography{refs}
\end{document}